\documentclass[10pt,aps,prb,twocolumn,showpacs,amssymb]{revtex4-1}
\usepackage{amsmath,graphics,epsfig,mathrsfs}
\usepackage{epstopdf}
\usepackage{bm}
\usepackage{color}
\usepackage{textcomp}
\usepackage{hyperref}
\hypersetup{backref=true,pagebackref=true,hyperindex=true,colorlinks=true,breaklinks=true,urlcolor=
red,linkcolor=blue,bookmarks=true,bookmarksopen=true,citecolor=red}

\begin{document}
\title{Spin Hall magnetoresistance in antiferromagnet/normal metal bilayers}
\author{Aur\'elien Manchon}
\affiliation{King Abdullah University of Science and Technology (KAUST), Physical Science and Engineering Division (PSE), Thuwal 23955-6900, Kingdom of Saudi Arabia}

\begin{abstract}
We predict the emergence of spin Hall magnetoresistance in a magnetic bilayer composed of a normal metal adjacent to an antiferromagnet. Based on a recently derived drift-diffusion equation, we show that the resistance of the bilayer depends on the relative angle between the direction transverse to the current flow and the N\'eel order parameter. Although this effect presents striking similarities with the spin Hall magnetoresistance recently reported in ferromagnetic bilayers, in the present case its physical origin is attributed to the anisotropic spin relaxation of itinerant spins in the antiferromagnet.
\end{abstract}
\maketitle

Transition metal multilayers have received a renewed interest lately with the search for current-driven spin-orbit torques \cite{Brataas2014,Manchon2014,Miron2011,Liu2012} and thermally-driven spin transport \cite{Bauer2012} in these systems. While bulk transition metal ferromagnets possess an anomalous conductivity tensor - hence displaying anisotropic magnetoresistance \cite{McGuire1975} and anomalous Hall transport \cite{Nagaosa2010} -, it has been recently realized that ultrathin films also display a peculiar form of the conductivity tensor. In particular, it has been shown that multilayers involving heavy metals possess a sizable anisotropic magnetoresistance with symmetries different from the one traditionally found in bulk ferromagnets \cite{Kobs2011}. While anisotropic magnetoresistance in bulk polycrystalline films \cite{McGuire1975} depends on the angle between the flowing current ${\bf j}_c$ and the magnetization direction, ${\bf m}$, i.e. $\sim ({\bf m}\cdot{\bf j}_c)^2$, in ultrathin films an additional (interfacial) anisotropic magnetoresistance emerges that depends on the angle between the magnetization and the direction {\em transverse} to the current flow, $\sim [{\bf m}\cdot({\bf z}\times{\bf j}_c)]^2$, where ${\bf z}$ is the normal to the multilayer interfaces. Various origins have been proposed to explain this effect, such as anisotropic spin scattering arising from semiclassical size effect \cite{Kobs2011,Kobs2012}, interfacial Rashba spin-orbit coupling \cite{Wang2014a,Zhang2015} and spin Hall effect taking place in the normal metal adjacent to the ferromagnet \cite{Chen2013,Nakayama2013}. Now confirmed in a wide range of transition metal magnetic bilayers \cite{Isasa2014,Kim2016,Vlietstra2013}, this effect is usually designated under the broad name of "spin Hall magnetoresistance" (SMR).\par

The research reported to date on the transport properties of ferromagnets has recently been extended to antiferromagnets, where spin-orbit torques \cite{Zelezny2014,Wadley2016} and spin Seebeck effect \cite{Wu2016,Prakash2016,Seki2016} have been explored. The field of antiferromagnetic spintronics is now blooming, bearing promises for potential spin-based devices\cite{Jungwirth2016,Baltz2016}. In his Nobel lecture, N\'eel stated that any properties of ferromagnets that are {\em even} under magnetization reversal should also exist in antiferromagnets \cite{Neel1970}. As a matter of fact, bulk anisotropic magnetoresistance \cite{Marti2014,Wang2014b,Fina2014}, as well as tunneling anisotropic magnetoresistance \cite{Park2011,Marti2012,Wang2013} have been observed in several metallic antiferromagnets already and signatures of spin-orbit torques have been reported in antiferromagnetic bilayers \cite{Reichlova2015,Yang2016}. In this work, using a recently derived drift-diffusion model\cite{Manchon2016}, we demonstrate that metallic bilayers composed of an antiferromagnet adjacent to a normal metal also exhibit spin Hall magnetoresistance, in a similar manner as their ferromagnetic counterparts.

\begin{figure}[h]
\centering
\includegraphics[width=5cm]{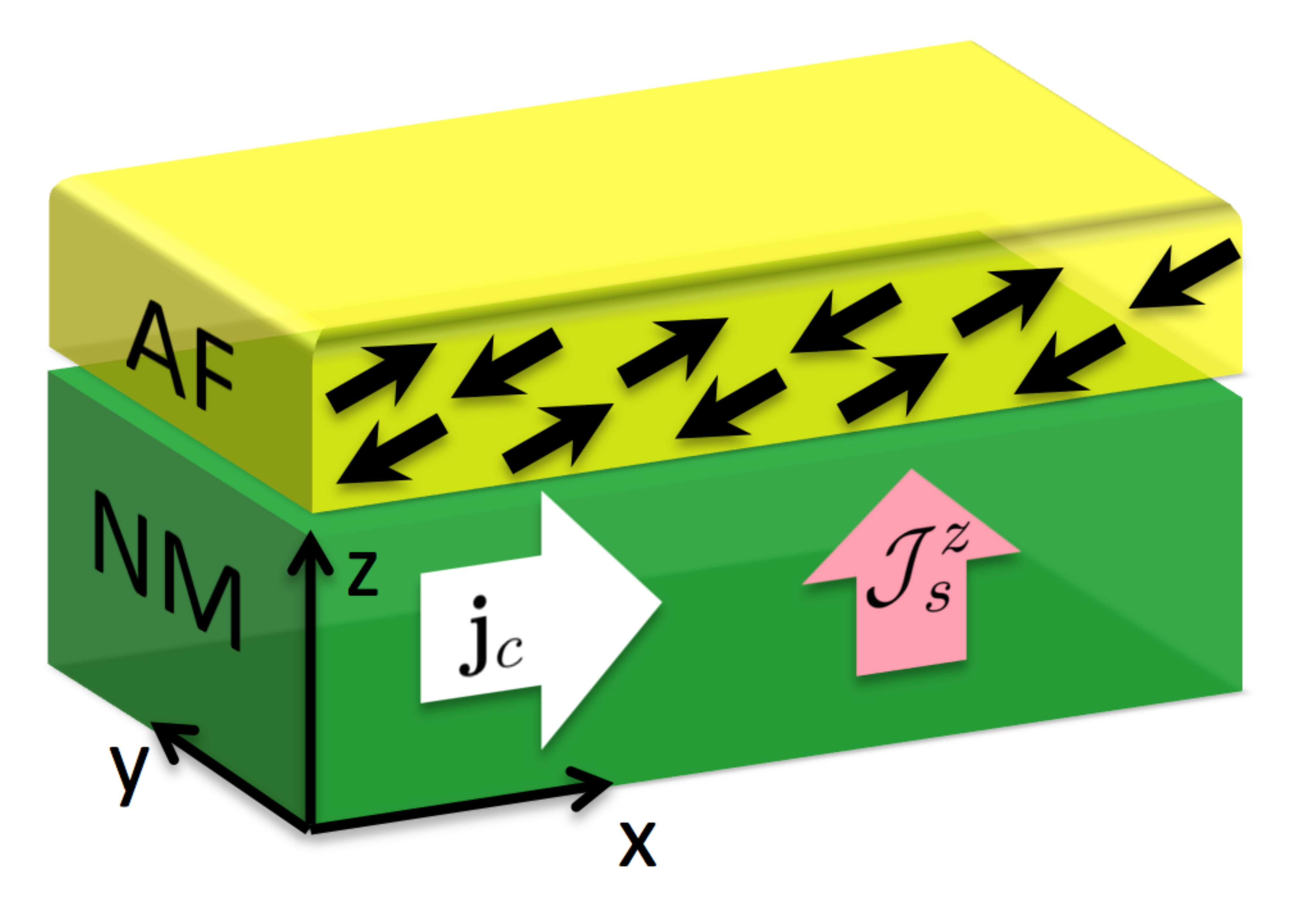}
\caption{(Color online) Schematics of the antiferromagnetic bilayer. The antiferromagnet (AF - yellow) is adjacent to a normal metal (NM - green). Due to spin-orbit coupling inside the normal metal, a flowing charge current along ${\bf x}$ (thick white arrow) creates a spin current flowing along ${\bf z}$ and polarized along ${\bf y}$ (thick pink arrow). The small black arrows represent the magnetic moments in the antiferromagnet.}
\label{Fig1}
\end{figure}

The system we consider is depicted in Fig. \ref{Fig1}. A collinear, bipartite antiferromagnet (yellow) is deposited on top of a normal metal (green). The current is injected along {\bf x} and the interface is normal to {\bf z}. The antiferromagnet possesses a G-type (checkerboard) magnetic configuration, characterized by its N\'eel order parameter {\bf n}. The normal metal possesses spin-orbit coupling so that a spin Hall effect emerges: a flowing charge current ${\bf j}_c$ creates a spin current ${\cal J}_s^z=(\theta_{\rm sh}/e){\bf e}_i\times{\bf j}_c$, where ${\cal J}_s^i$ is the $i$-th spatial component of the spin current and $\theta_{\rm sh}$ is the spin Hall angle. Notice that ${\cal J}_s$ is a 3$\times3$ tensor, while ${\bf j}_c$ is a three dimensional vector. The spin diffusion equations in the normal metal read\cite{Shchelushkin2005,Haney2013}
\begin{eqnarray}
&&-e\partial_i{\cal J}_s^i/{\cal N}=\frac{1}{\tau_{\rm sf}^{\rm N}}{\bm\mu},\\
&&e{\cal J}_s^i=-\sigma_{\rm N}\partial_i{\bm\mu}+\theta_{\rm sh}^{\rm N}({\bf e}_i\times{\bm \nabla})\mu_c.\label{eq:Jsn}
\end{eqnarray}
Here $\mu_c$ (${\bm \mu}$) is the scalar (vector) of spin-dependent (spin-independent) electrochemical potential, ${\cal N}$ is the density of states and $\sigma_{\rm N}$ is the conductivity in the normal metal.\par

In a recent work, we derived the drift-diffusion equation for collinear, bipartite antiferromagnets based on quantum kinetic principles \cite{Manchon2016}. In this model, the metallic antiferromagnet is composed of two magnetic sublattices, say A and B, aligned antiferromagnetically with each other. The spin-dependent electrochemical potential on each sublattice can we written ${\bm\mu}_{A,B}={\bm \mu}\pm\delta{\bm\mu}$, where ${\bm \mu}$ is the {\em uniform} component averaged over the unit cell, while $\delta{\bm\mu}$ is the {\em staggered} component. This staggered component, only present in the antiferromagnet, is crucial to obtain spin torque \cite{Manchon2016} but does not play any role for the effect discussed in the present work. Interestingly, in the antiferromagnet the uniform spin-dependent electrochemical potential fulfills the following drift-diffusion equation
\begin{eqnarray}\label{eq:AFs}
&&-e\partial_i{\cal J}_s^i/{\cal N}=\frac{1}{\tau_\varphi^{\rm AF}}{\bf n}\times({\bm\mu}\times{\bf n})+\frac{1}{\tau_{\rm sf}^{\rm AF}}{\bm\mu},
\end{eqnarray}
where the first term in the right-hand side, $\sim 1/\tau_{\varphi}^{\rm AF}$, amounts for the spin dephasing that relaxes only the spin component that is {\em transverse} to the N\'eel order parameter, while the second term, $\sim 1/\tau_{\rm sf}^{\rm AF}$, is the conventional isotropic spin relaxation (driven by spin-orbit coupling, magnetic impurities etc.). Equation (\ref{eq:AFs}) resembles the spin diffusion equation in ferromagnets, except that the spin precession term is absent. The spin current reads
\begin{eqnarray}
&&e{\cal J}_s^i=-\sigma_\|^{\rm AF}(\partial_i{\bm\mu}\cdot{\bf n}){\bf n}-\sigma_\bot^{\rm AF}{\bf n}\times(\partial_i{\bm\mu}\times{\bf n}),
\end{eqnarray}
where $\sigma_{\|(\bot)}^{\rm AF}$ is the conductivity of carriers whose spin lies along (perpendicular to) the N\'eel order parameter ${\bf n}$. In other words, the drift-diffusion equation that governs the transport of the uniform spin-dependent electrochemical potential ${\bm \mu}$ in antiferromagnets is quite similar to that in a normal metal, except it is {\em anisotropic} with respect to the N\'eel order parameter {\bf n}.\par

We now compute the spin-dependent electrochemical potential profile in the antiferromagnetic bilayer depicted in Fig. \ref{Fig1}. For the boundary conditions, we neglect interfacial spin-flip so that ${\cal J}_s^z|_{z=0^-}={\cal J}_s^z|_{z=0^+}$. Notice that interfacial spin-flip can be added by hand but renders the analytical expressions cumbersome \cite{Liu2014}. Since the spin transport in the antiferromagnet is anisotropic, we also consider an anisotropic interfacial resistivity such that 
\begin{eqnarray}
&&({\bm\mu}_{\rm AF}-{\bm\mu}_{\rm N})\cdot{\bf n}=r_\|{\cal J}_s^z\cdot{\bf n},\\
&&{\bf n}\times[({\bm\mu}_{\rm AF}-{\bm\mu}_{\rm N})\times{\bf n}]=r_\bot{\bf n}\times({\cal J}_s^z\times{\bf n}),
\end{eqnarray}
where $r_{\|(\bot)}$ is the interfacial resistivity for spins parallel (transverse) to the N\'eel order parameter. No spin current flows through the outer boundaries, ${\cal J}_s^z|_{z=-d_{\rm N}}={\cal J}_s^z|_{z=d_{\rm AF}}=0$, where $d_{\rm N}$ and $d_{\rm AF}$ denote the thickness of the normal metal and antiferromagnetic layers, respectively. We obtain the spin-dependent electrochemical potential in the structure
\begin{widetext}
\begin{eqnarray}\label{eq:muAF}
\frac{{\bm\mu}_{\rm AF}}{\nabla_x\mu_c}&=&-\left[\frac{\gamma_\bot\theta_{\rm sh}\lambda_{\rm sf}^{\rm N}}{1+\gamma_\bot\eta_\bot\tanh\frac{d_{\rm N}}{\lambda_{\rm sf}^{\rm N}}}\frac{\cosh\frac{z-d_{\rm AF}}{\lambda_\bot^{\rm AF}}}{\cosh\frac{d_{\rm AF}}{\lambda_\bot^{\rm AF}}}{\bf n}\times({\bf y}\times{\bf n})+\frac{\gamma_\|\theta_{\rm sh}\lambda_{\rm sf}^{\rm N}}{1+\gamma_\|\eta_\|\tanh\frac{d_{\rm N}}{\lambda_{\rm sf}^{\rm N}}}\frac{\cosh\frac{z-d_{\rm AF}}{\lambda_\|^{\rm AF}}}{\cosh\frac{d_{\rm AF}}{\lambda_\|^{\rm AF}}}n_y{\bf n}\right]\left(1-\cosh^{-1}\frac{d_{\rm N}}{\lambda_{\rm sf}^{\rm N}}\right),\\
\frac{{\bm\mu}_{\rm N}}{\nabla_x\mu_c}&=&-\left[\frac{\sinh\frac{z}{\lambda_{\rm sf}^{\rm N}}+\gamma_\bot\eta_\bot\left(\cosh\frac{z+d}{\lambda_{\rm sf}^{\rm N}}-\cosh\frac{z}{\lambda_{\rm sf}^{\rm N}}\right)}{1+\gamma_\bot\eta_\bot\tanh\frac{d_{\rm N}}{\lambda_{\rm sf}^{\rm N}}}{\bf n}\times({\bf y}\times{\bf n})+\frac{\sinh\frac{z}{\lambda_{\rm sf}^{\rm N}}+\gamma_\|\eta_\|\left(\cosh\frac{z+d}{\lambda_{\rm sf}^{\rm N}}-\cosh\frac{z}{\lambda_{\rm sf}^{\rm N}}\right)}{1+\gamma_\|\eta_\|\tanh\frac{d_{\rm N}}{\lambda_{\rm sf}^{\rm N}}}n_y{\bf n}\right]\frac{\theta_{\rm sh}\lambda_{\rm sf}^{\rm N}}{\cosh\frac{d_{\rm N}}{\lambda_{\rm sf}^{\rm N}}}.\nonumber\label{eq:muN}\\
\end{eqnarray}
\end{widetext}
where 
\begin{eqnarray}
&&\eta_\alpha=1+(r_\alpha\sigma_\alpha^{\rm AF}/\lambda_\alpha^{\rm AF})\tanh\frac{d_{\rm AF}}{\lambda_\alpha^{\rm AF}},\\
&&\gamma_\alpha=(\lambda_\alpha^{\rm AF}\sigma_{\rm N}/\lambda_{\rm sf}^{\rm N}\sigma_\alpha)\tanh^{-1}\frac{d_{\rm AF}}{\lambda_\alpha^{\rm AF}},
\end{eqnarray} 
with $\alpha=\|,\bot$. We defined
\begin{eqnarray}
\lambda_\|^{\rm AF}&=&\sqrt{{\cal D}_\|^{\rm AF}\tau_{\rm sf}^{\rm AF}},\\
\lambda_\bot^{\rm AF}&=&\sqrt{{\cal D}_\bot^{\rm AF}/(1/\tau_{\rm sf}^{\rm AF}+1/\tau_\varphi^{\rm AF})},
\end{eqnarray}
where ${\cal D}_{\|(\bot)}^{\rm AF}=\sigma_{\|(\bot)}^{\rm AF}/e^2{\cal N}$. Assuming that ${\cal D}_\|^{\rm AF}={\cal D}_\bot^{\rm AF}$, we obtain $\lambda_\|^{\rm AF}/\lambda_\bot^{\rm AF}=\sqrt{1+\tau_{\rm sf}^{\rm AF}/\tau_\varphi^{\rm AF}}$: the stronger the spin dephasing, the larger the anisotropy of the spin relaxation. Finally, the charge current flowing through the normal metal reads \cite{Shchelushkin2005,Haney2013}
\begin{eqnarray}\label{eq:je}
{\bf j}_c=-\sigma_{\rm N}\partial_x\mu_c-\theta_{\rm sh}\sigma_{\rm N}\partial_x\mu_y,
\end{eqnarray} 
and therefore, the spin Hall magnetoresistance is given by the change in $\mu_y$ profile when the N\'eel order changes direction. \par

Let us now compute the spin-dependent electrochemical potential and the associated spin Hall magnetoresistance. For simplicity and in the absence of detailed experimental data, we neglect the anisotropy in both conductivity and interfacial resistivity and only consider the impact of spin dephasing in the antiferromagnet. We choose ${\cal D}_\|^{\rm AF}={\cal D}_\bot^{\rm AF}=0.4\times10^{-3}$ m$^2\cdot$s$^{-1}$, $\sigma_{\rm AF}^\|=\sigma_{\rm AF}^\bot=\sigma_{\rm N}=10^6$ $\Omega^{-1}\cdot$m$^{-1}$, and $r_\|=r_\bot=0.3\;{\rm m}\Omega\cdot\mu$m$^2$. Finally, for the spin relaxation and dephasing times, we choose $\tau_{\rm sf}^{\rm N}=6\times10^{-14}$ s, $\tau_{\rm sf}^{\rm AF}=10^{-14}$ s and $\tau_{\varphi}^{\rm AF}=10^{-15}$ s, such that $\lambda_{\rm sf}^{\rm N}=5$ nm (unless stated otherwise), $\lambda_\|^{\rm AF}=2$ nm and $\lambda_\bot^{\rm AF}=0.6$ nm. The magnitude of the spin relaxation and dephasing lengths are consistent with recent experimental reports \cite{Merodio2014,Zhang2015}. 

\begin{figure}[h]
\centering
\includegraphics[width=8cm]{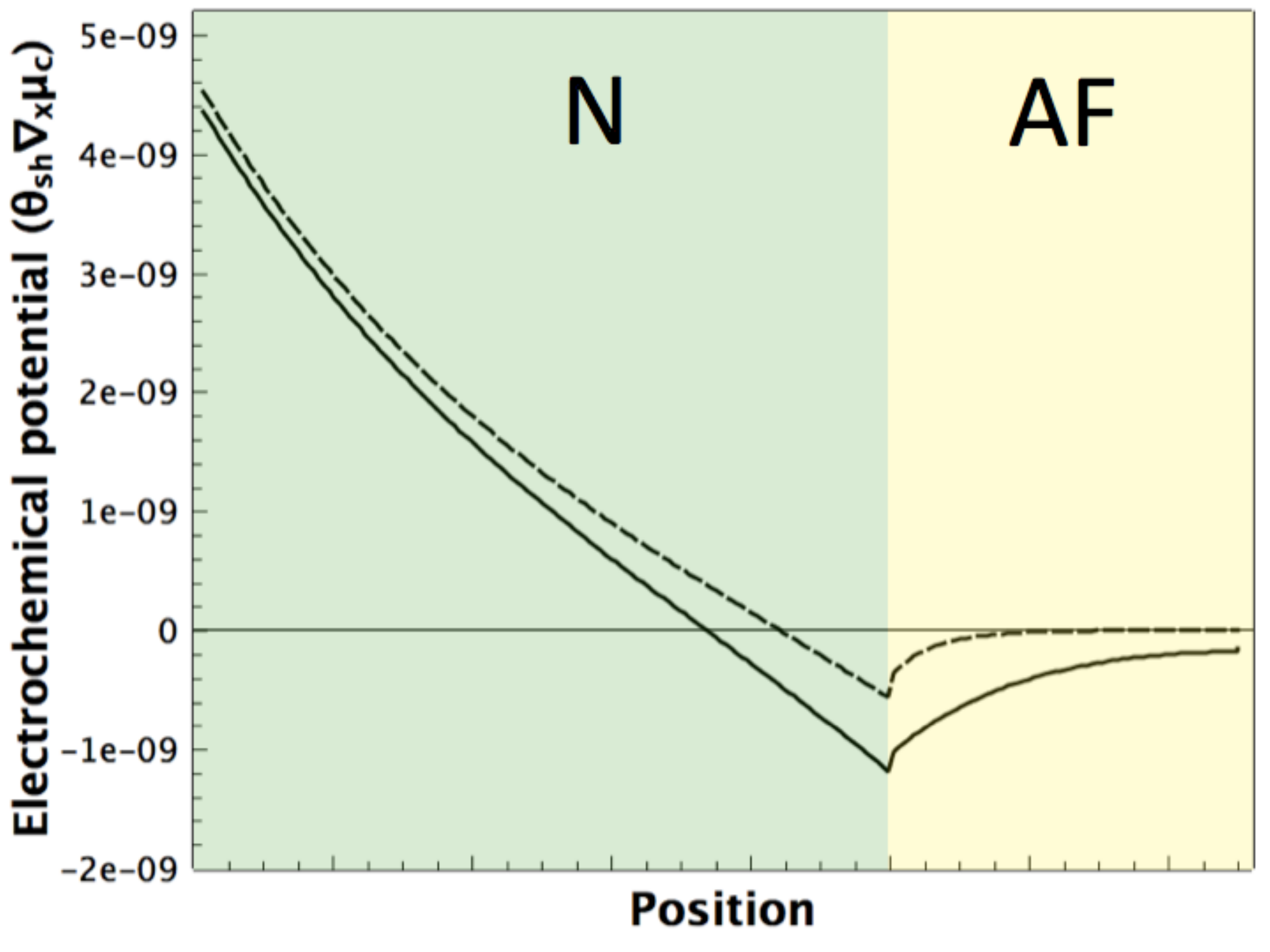}
\caption{(Color online) Spatial profile of the spin-dependent electrochemical potential $\mu_y$ when the N\'eel order parameter lies normal (solid line) and along (dashed line) to the current direction. In these calculations, $d_{\rm N}=10$ nm, and $d_{\rm AF}=5$ nm. The other parameters are given in the text.}
\label{Fig2}
\end{figure}

The spatial profile of the spin-dependent electrochemical potential $\mu_y$ through the structure is given in Fig. \ref{Fig2} for two configurations of the magnetic order, $n_y=1$ (solid line) and $n_y=0$ (dashed line). In the normal metal (N), $\mu_y$ is driven by spin Hall effect, while in the antiferromagnet (AF) $\mu_y$ simply relaxes with different decay rates depending on the direction of the N\'eel order parameter. Therefore, the change in spin-dependent electrochemical potential is associated with the anisotropy of the spin relaxation length characteristic of collinear antiferromagnets. 

Injecting Eq. (\ref{eq:muN}) into Eq. (\ref{eq:je}), one obtains the expression of the spin Hall magnetoresistance,
\begin{eqnarray}\label{eq:smr}
\Delta\sigma_{xx}=\frac{(\gamma_\|\eta_\|-\gamma_\bot\eta_\bot)(1-\cosh^{-1}\frac{d_{\rm N}}{\lambda_{\rm sf}^{\rm N}})^2\lambda_{\rm sf}^{\rm N}\sigma_{\rm N}\theta_{\rm sh}^2}{(1+\gamma_\|\eta_\|\tanh^{-1}\frac{d_{\rm N}}{\lambda_{\rm sf}^{\rm N}})(1+\gamma_\bot\eta_\bot\tanh^{-1}\frac{d_{\rm N}}{\lambda_{\rm sf}^{\rm N}})}.\nonumber\\
\end{eqnarray}
The longitudinal conductivity is simply $\sigma_{xx}\approx d_{\rm N}\sigma_{\rm N}+d_{\rm AF}\sigma_{\rm AF}+O(\theta_{\rm sh}^2)$. More specifically, the spin Hall magnetoresistance is proportional to the anisotropy of the spin transport in the antiferromagnet,
\begin{eqnarray}
\frac{\Delta\sigma_{xx}}{\sigma_{xx}}\sim (\frac{\lambda_\|^{\rm AF}}{\sigma_\|}-\frac{\lambda_\bot^{\rm AF}}{\sigma_\bot})+(r_\|\tanh\frac{d_{\rm AF}}{\lambda_\bot^{\rm AF}}-r_\bot\tanh\frac{d_{\rm AF}}{\lambda_\bot^{\rm AF}}).\nonumber\\
\end{eqnarray}
Equation \eqref{eq:smr} presents striking similarities with the one derived in ferromagnetic bilayers \cite{Kim2016}. In the language of the mixing conductance \cite{Brataas2000}, one can identify the real part of the interfacial spin mixing conductance, $2{\rm Re}G^{\uparrow\downarrow}=\sigma_\bot^{\rm AF}\tanh\frac{d_{\rm AF}}{\lambda_\bot^{\rm AF}}/\lambda_\bot^{\rm AF}$. This relation is revealing as the spin Hall magnetoresistance in ferromagnetic and antiferromagnetic bilayers arise from the same source, i.e. the different interfacial spin resistance when the (ferro or antiferro) magnetic order lies along the direction ${\bf z}\times{\bf j}_c$ or normal to it. Recent theories have computed the spin mixing conductance for special cases of antiferromagnets \cite{Cheng2014,Takei2014}. In our theory, this spin mixing conductance is associated with the transverse spin dephasing in antiferromagnets. With our set of parameters, we obtain $2{\rm Re}G^{\uparrow\downarrow}\approx 1.7\times 10^{15}$ $\Omega^{-1}\cdot$ m$^{-2}$ (for $d_{\rm AF}\gg\lambda_{\bot}^{\rm AF}$), a value comparable to that of ferromagnets.\par

\begin{figure}[h]
\centering
\includegraphics[width=8cm]{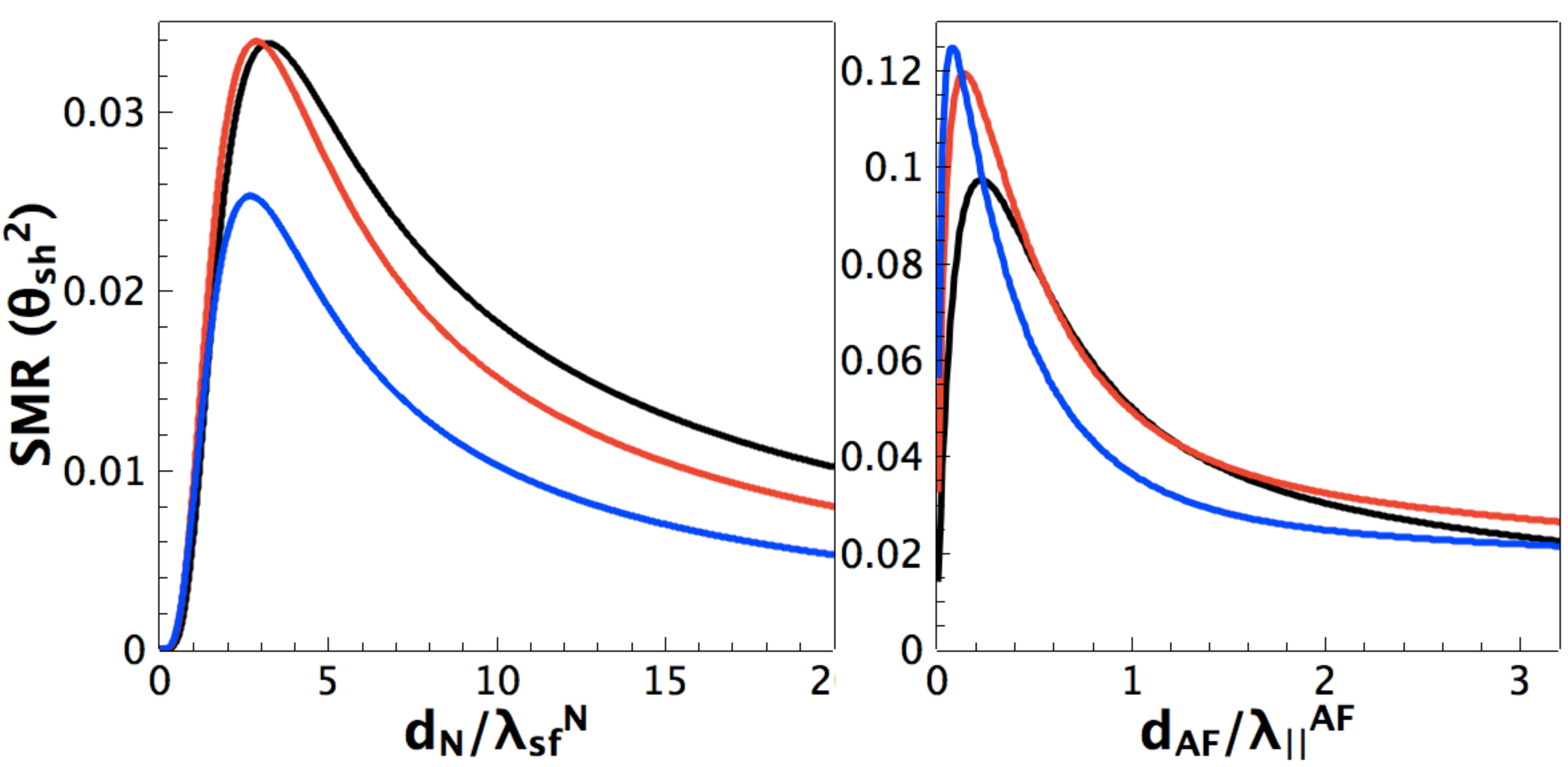}
\caption{(Color online) Spin Hall magnetoresistance ratio as a function of the thickness of (a) the normal metal (with $d_{\rm AF}=5$ nm) and (b) the antiferromagnet (with $d_{\rm N}=10$ nm). In these calculations, the spin relaxation length of the normal metal has been taken at $\lambda_{\rm sf}^{\rm N}=2$ (black), 5 (red) and 10 nm (blue), respectively.}
\label{Fig3}
\end{figure}

Figure \ref{Fig3}(a,b) represents the spin Hall magnetoresistance as a function of the thickness of (a) the normal metal and (b) the antiferromagnet. The dependence as a function of the normal metal thickness shows a peak, which reveals a competition between the progressive build-up of the spin Hall effect in the normal metal (for $d_{\rm N}\lesssim\lambda_{\rm sf}^{\rm N}$) and the shunting of the current (for $d_{\rm N}>\lambda_{\rm sf}^{\rm N}$). The dependence as a function of the antiferromagnet thickness shows a similar behavior with a sharp increase at small thicknesses, corresponding to the quenching of the transverse spin-dependent electrochemical potential in the antiferromagnet (for $d_{\rm AF}\lesssim\lambda_{\bot}^{\rm AF}$), and a slow decay corresponding to the shunting of the current through the antiferromagnet (for $d_{\rm AF}>\lambda_{\bot}^{\rm AF}$).\par

The spin Hall magnetoresistance predicted in this work should be observable in any multilayers involving collinear antiferromagnets. The experimental observation of this effect requires manipulating the N\'eel order parameter, which can be achieved using exchange bias with a proximate ferromagnet \cite{Fina2014}, field-cooling procedure \cite{Marti2014,Moriyama2015a}, or spin-orbit torque in the case of non-centrosymmetric antiferromagnets \cite{Wadley2016}. Noticeably, this effect is not limited to metals and should be observable in multilayers comprising insulating antiferromagnets such as NiO, CoO, Cr$_2$O$_3$ etc. In this case, the interfacial mixing conductance is associated with the absorption the spin current mediated by (coherent or incoherent) spin waves inside the antiferromagnetic insulator \cite{Takei2014,Takei2015}. Recent experiments suggest that the absorption length can be quite large \cite{Wang2014c,Hahn2014,Moriyama2015b} (Wang et al. have reported a decay length of 10 nm in NiO \cite{Wang2014c}), and the associated mixing conductance can be much larger than their ferromagnetic counterpart due to its high temperature sensitivity \cite{Lin2016,Seki2015}. Finally, an interesting question that remains to be addressed is whether spin Hall magnetoresistance could be observed in {\em non-collinear} antiferromagnets. A simple-minded argument suggests that as long as a magnetic order parameter can be defined, spin Hall magnetoresistance should emerge, respecting the symmetries of the antiferromagnetic overlayer. Nevertheless, a thorough investigation of realistic systems is necessary.

In conclusion, we have showed that bilayers composed of a collinear antiferromagnet adjacent to a normal metal with spin-orbit coupling should exhibit spin Hall magnetoresistance, similar to their ferromagnetic counterpart. In our model, the mechanism responsible for this effect is the anisotropic relaxation of itinerant spins inside the antiferromagnet with respect to the N\'eel order parameter. Several experimental methods have been recently used to investigate the emergence of anisotropic magnetoresistance in bulk collinear antiferromagnets, indicating viable routes for the detection of antiferromagnetic spin Hall magnetoresistance.

\acknowledgments
We acknowledge the financial support of the King Abdullah University of Science and Technology (KAUST) through the Office of
Sponsored Research (OSR) [Grant Number OSR-2015-CRG4-2626].

\end{document}